\documentclass{cpbtex}

\usepackage{graphicx}
\usepackage{subcaption}
\usepackage{footnote}
\usepackage{caption}
\usepackage{dcolumn}
\usepackage{bm}
\usepackage[inline]{enumitem}
\usepackage{color, colortbl}
\usepackage[svgnames]{xcolor}
\usepackage{multirow}
\usepackage{soul}
\usepackage{wasysym}

\newcommand{\myeqref}[1]{Eq.~\eqref{#1}}

\newcommand{\mycite}[1]{\ucite{#1}}
\newcommand{\myRefInd}[1]{Ref.~\cite{#1}}
\newcommand{\mytabref}[1]{Table.~\ref{#1}}

\newcommand{\Rmnum}[1]{\uppercase\expandafter{\romannumeral #1}}

\newcommand{\rvs}[1] { #1 }
\newcommand{\mrvs}[1] { #1 }

\begin{document}

\title{On-node lattices construction using \textit{partial} Gauss-Hermite quadrature for the lattice Boltzmann method \thanks{Project supported by National Science and Technology Major Project (Grant No.~2017ZX06002002).} }

\author{Huanfeng Ye$^{1}$\thanks{Corresponding author. E-mail: huanfye@163.com},
	\ Zecheng Gan$^{2}$, 
	\ Bo Kuang$^{1}$ 
	\ and \ Yanhua Yang$^{1,3}$\\
	$^{1}${School of Nuclear Science and Engineering, Shanghai Jiao Tong University} \\ {Shanghai 200240, China}\\
	$^{2}${Department of Mathematics, University of Michigan}\\
	{Ann Arbor, MI 48109-1043, USA}\\
	$^{3}${National Energy Key Laboratory of Nuclear Power Software }\\
	{Beijing 102209, China}}

\date{\today}
\maketitle

\begin{abstract}
 A concise theoretical framework, the \textit{partial} Gauss-Hermite quadrature (pGHQ), is established for constructing on-node lattices of the lattice Boltzmann (LB) method under a Cartesian coordinate system. Comparing with \mrvs{existing} approaches, the pGHQ scheme has the following advantages: \textbf{a).}  extremely concise algorithm, \textbf{b).} unifying the constructing procedure of symmetric and asymmetric on-node lattices, \textbf{c).} covering full-range quadrature degree of a given discrete velocity set. We employ it to search the local optimal and asymmetric lattices 
for $\left\{ {n = 3,4,5,6,7} \right\}$ moment degree equilibrium distribution discretization on range $\left[ { - 10,10} \right]$. The search reveals a surprising abundance of available lattices. Through a brief analysis, the discrete velocity set shows a significant influence on the positivity of equilibrium distributions, \mrvs{which is considered as} one major impact to the numerical stability of the LB method. 
Hence the results of the pGHQ scheme lay a foundation for further investigations on improving the numerical stability of the LB method by modifying the discrete velocity set.
It also worths noting that pGHQ can be extended into the entropic LB model though it was proposed for the Hermite polynomial expansion LB theory.
\end{abstract}



\textit{Introduction.--} The lattice Boltzmann (LB) method is a powerful approach for hydrodynamics \cite{DorschnerBösch-2390,FrapolliChikatamarla-2396}. The essence of the LB method is an intuitively parallel collision-streaming algorithm with discretized position $\vec r$, time $t$ and microscopic velocity $\vec{v}_\alpha$,
\begin{equation}
\label{eq:LB}
{f_\alpha }(\vec r + {{\vec v}_\alpha {\delta_t}, t + {\delta_t} }) = \left( {1 - \frac{1}{\tau }} \right){f_\alpha }(\vec r, t) + \frac{1}{\tau }f_\alpha ^{eq}(\vec r, t)
\end{equation}
where $f_\alpha$ and $f_\alpha^{eq}$ are, respectively, the population and equilibrium distribution corresponding to the discrete velocity ${\vec v}_\alpha$.  
\myeqref{eq:LB} can be treated as a characteristic integral of the BGK-Boltzmann equation along ${\vec v}_\alpha$ \cite{HeLuo-2016,HeChen-173}, depicting the microscopic dynamic of particles. With specific discretization of the continuous  BGK-Boltzmann equation in velocity space, i.e. on-node lattices, each collision-streaming proceeding would locate on nodes, achieving simple but efficient ``stream along links and collide at nodes'' algorithm, meanwhile the corresponding macroscopic dynamics such as the Navier-Stokes equations can be properly recovered. In practice, this velocity discretization can be achieved through constructing a set of equilibrium distributions $\{f_\alpha^{eq}\}$ on a discrete velocity set $\{v_\alpha\}$, i.e. equilibrium distribution (ED) discretization. Under a Cartesian coordinate system, multidimensional $f_\alpha^{eq}$ can always be constructed as a tensor product of the unidimensional one so that we will focus on the unidimensional Cartesian model to simplify our framework.

ED discretization has been in-depth investigated, and a lot of excellent theories have been proposed, e.g. the small-Mach-number approximation\mycite{FrischD'Humieres-2782}, the Hermite polynomial expansion\mycite{ShanHe-2080} and the entropic LB model\mycite{KarlinFerrante-2508}. According to \mrvs{the Hermite polynomial expansion}\mycite{Shan-2088,Shan-2582,ShanHe-2080},  for $n$th-moment-order ED discretization which restores $u^n$ moment integral, its $f_\alpha^{eq}$ can be expressed as
\begin{equation}
\label{eq:disEq}
f_\alpha ^{eq} = {{\text w}_\alpha }\rho \sum\limits_{i = 0}^n {{H_i}\left( \xi_\alpha  \right)\frac{{{\phi ^i}}}{{i!}}} ,
\end{equation}
with
\begin{align}
\label{eq:mom}
\rho  &= \sum\limits_{\alpha  = 0}^{q - 1} {f_\alpha ^{eq}} , \quad
\phi  = \sum\limits_{\alpha  = 0}^{q - 1} {f_\alpha ^{eq}{\xi _\alpha }} /\rho ,\\
\label{eq:wght}
{{\text w}_\alpha } &= \frac{1}{{\sqrt \pi  }}\int {e^{ { - {\xi ^2}} }\prod\limits_{\substack{ 
			\beta  = 0 \\
			\beta  \ne \alpha}} ^{q-1} {\frac{{\xi  - {\xi _\beta }}}{{{\xi _\alpha } - {\xi _\beta }}}} d\xi } .
\end{align}
where $\xi = v/\sqrt{2RT}$ and $\phi = u/\sqrt{2RT}$ are the dimensionless variables of  microscopic velocity $v$ and macroscopic velocity $u$ respectively in which $R$ is the gas constant and $T$ the temperature, $H_i\left(\xi\right)$ is the $i$th Hermite polynomial,
and $\left\{ {\xi _0}, \ldots ,{\xi _{q-1}}\right\}$ is the corresponding discrete set (or abscissas) which evaluates the integral exactly for $k\le 2n$,
\begin{equation}
\label{eq:qEq}
I^k=\int {\frac{1}{{\sqrt {\pi} }}e^ { { - {\xi ^2}} }{\xi ^k}d\xi }  = \sum\limits_{\alpha  = 0}^{q - 1} {{{\text w}_\alpha }\xi _\alpha ^k}.
\end{equation} 
\mrvs{The abscissas and the discrete velocity set have the relation $\xi_\alpha=v_\alpha /\sqrt{2RT}$.}
The Hermite polynomial expansion converts the $n$th-moment-order ED discretization under a unidimensional Cartesian coordinate system into a pure $2n$-degree quadrature problem, i.e. constructing the smallest abscissas $\left\{ {\xi _0}, \ldots ,{\xi _{q-1}}\right\}$ fulfilling \myeqref{eq:qEq} for all $k\le 2n$. One can refer to Sec.~\Rmnum{1} in the supporting information for the detail derivation.  For the sake of simplifying discussion, we designate \myeqref{eq:qEq}  as quadrature equation (QE) and  its equation system of all $k \le n$ as $n$th quadrature equation system (QES). It should be noted that $n$th QES is the detail governing equation system for a given abscissa set $\left\{ \xi_\alpha \right\}$ with quadrature degree $n$. Hence in the discussion hereinafter, {\it QES} and {\it quadrature degree} shall be used indistinguishably.

An available smallest quadrature for $2n$th QES is the $\left(n+1\right)$th Gauss-Hermite quadrature, \mrvs{which} are the zeros of $\left(n+1\right)$th Hermite polynomial. The issue is that the zeros of an Hermite polynomial with degree above 3 cannot fit into nodes, which means that they can not be expressed as $\{v_0,\ldots,v_{q-1}\}c$ where $v$ and $c$ stand for integer-valued discrete micro velocity and real-valued lattice constant $1/\sqrt{2RT}$ respectively. It leads to an off-node lattices in $u^{n\ge3}$ ED discretization. 
Hence to construct an on-node lattices for $u^{n\ge3}$ ED discretization, \mrvs{i.e.}  $\{v_0,\ldots,v_{q-1}\}c$-type quadrature, one has to manually solve QES which involves both $\left\{ \xi_\alpha\right\}$ and $\left\{\text{w}_\alpha\right\}$. 
To simplify the notation, in the discussion hereinafter, \textit{lattices} would directly denote \textit{on-node lattices} unless otherwise stated.
In practice, a symmetric discrete velocity set $\{0, \pm v_1,\ldots,\pm v_{m}\}$ is predefined. It avoids the computation of QE with odd exponent $k$ which significantly simplifies QES, and makes QES purely consist of $c$ and $\left\{\text{w}_\alpha\right\}$. Employing the skills in \myRefInd{Shim-2780,Shan-2582} to deal with QES, an univariate polynomial equation for lattice constant $c$ can be obtained, \mrvs{which separates} the co-solving of $c$ and $\left\{\text{w}_\alpha\right\}$. It leads us to a performable construction of on-node lattices.
Actually, this univariate polynomial equation can be directly obtained through a mathematical tool avoiding the tedious QES solving.

In this paper, this mathematical tool, the \textit{partial} Gauss-Hermite quadrature (pGHQ), is proposed.
The tool name is emphasized by adding an italic adjective to distinguish from its origin and reflect the characteristic. 
pGHQ is a quadrature rule derived from the Gauss-Hermite quadrature. \mrvs{It keeps the most desirable characteristic of the Gauss-Hermite Quadrature, i.e. its quadrature is constructed directly on abscissa polynomial avoiding the co-solving of $\left\{ \xi_\alpha\right\}$ and $\left\{\text{w}_\alpha\right\}$ in QES.  Meanwhile it offers a performable approach for on-node lattices construction.} The on-node lattices construction in the pGHQ scheme is extremely concise. 
And once a discrete velocity set was given, a full-range univariate polynomial equation system of its lattice constant $c$ would be directly obtained through pGHQ. 
Comparing with the \mrvs{existing} schemes, our approach has the following advantages: \textbf{a).} the algorithm is extremely concise, \textbf{b).} the procedure of constructing univariate polynomial equations is unified for both symmetric and asymmetric lattices, \textbf{c).} the generated univariate polynomial equation system covers full-range quadrature degree of the given $\{v_0,\ldots,v_{q-1}\}$. We will elaborate them detailedly in the following.

\textit{pGHQ theory and implementation.--} The theory of pGHQ can be stated as:  for a $q$-point abscissa set $\{\xi_\alpha\}$, whose abscissa polynomial ${W_q}\left( \xi  \right) = \prod\limits_{\alpha  = 0}^{q - 1} {\left( {\xi  - {\xi _\alpha }} \right)} $ satisfies the orthogonal relationship
\begin{equation}
\label{eq:pGOth}
\int {e^{ { - {\xi ^2}} }{W_{q}}\left( \xi  \right)p\left( \xi  \right)d\xi }  = 0,\forall p\left( \xi  \right) \in {\mathbb{P}_{K\left( K<q  \right)}},
\end{equation}
where $\mathbb{P}_K$ is the set of polynomials of degrees not exceeding $K$, it has $\left(q+K\right)$ quadrature degree indicating that the set $\{\xi_\alpha\}$ and its corresponding $\{\text{w}_\alpha\}$ calculated by \myeqref{eq:wght} fulfills $\left(q+K\right)$th QES . pGHQ is a generalization of the Gauss-Hermite quadrature, \mrvs{which} is the special case of \myeqref{eq:pGOth} with polynomial degree $K=\left(q-1\right)$. \rvs{Given any polynomial $p\left( \xi  \right)$ of degree not exceeding $K$, it can always be expressed as a linear combination of Hermite polynomials with degree not exceeding $K$,
\begin{equation}\label{key}
p\left( \xi  \right) = \sum_{i=0}^{K}{c_i}{\xi^i}= \sum_{i=0}^{K}{a_i}{H_i\left( \xi  \right)},
\end{equation} 
Employing the orthogonal relationship of Hermite polynomials, 
\begin{equation}\label{key}
\int{{e^{ - {\xi ^2}}}{H_i}\left( \xi  \right){H_j}\left( \xi  \right)d\xi}  = \left\{ {\begin{array}{*{20}{c}}
	0&{i \ne j}\\
	{{2^i}i!\sqrt \pi}&{i = j}
	\end{array}} \right. ,
\end{equation}
the orthogonality in \myeqref{eq:pGOth} indicates that for a $q$-point quadrature with $q+K$ quadrature degree, its abscissa polynomial does not involve Hermite polynomials with degree below $K+1$ when written in Hermite polynomial form, i.e. all the coefficients of Hermite polynomials with degree below $K+1$ are zero,
\begin{equation}
\label{eq:weRal}
{W_q}\left( \xi  \right) = \sum\limits_{i = 0}^{q} {{A_i}{H_i}\left( \xi  \right)} = \sum\limits_{i = K + 1}^{q-1} {{A_i}{H_i}\left( \xi  \right)}  + \frac{1}{{{2^q}}}{H_q}\left( \xi  \right).
\end{equation}
As $A_i$ is an expression of the abscissas $\{\xi_\alpha\}$, then the zero coefficients  $\{A_i=0|i \le K\}$ could be used as the governing equation system of the 
abscissas under pGHQ for $q+K$ quadrature degree.  For the detail derivation, one can refer to Sec.~\Rmnum{2} in the supporting information. Hence for a $q$-point set $\{\xi_\alpha\}$,  once the coefficient equations $\{A_i=0\}$  were satisfied for all $i \le K$ in its Hermite-polynomial-form abscissa polynomial \myeqref{eq:weRal}, this set and its corresponding $\{\text{w}_\alpha\}$ in \myeqref{eq:wght} fulfills  $\left(q+K\right)$th QES. This coefficient equation system is denoted as Hermite coefficient equation system (HCES) in this paper.   
To identify a HCES, the denotation $q\sim  K^{th}$ is added before HCES, in which $q$ is the abscissa number, $K$ denotes the equations contained in the HCES $\{A_i=0|i \le K\}$, and $\left(q+K\right)$ is its corresponding quadrature degree. }
HCES is equivalent to QES but without involving $\left\{\text{w}_\alpha\right\}$. 
It indicates that pGHQ owns the desirable characteristic of the Gauss-Hermite Quadrature, constructing the quadrature directly on abscissa polynomial avoiding the co-solving of $\left\{ \xi_\alpha\right\}$ and $\left\{\text{w}_\alpha\right\}$ in QES.

Now we employ pGHQ to construct the univariate polynomial equation of $c$.
The univariate polynomial equation of $c$ essentially is a relation between $c$ and the quadrature degree of the corresponding abscissa set $\{v_0,\ldots,v_{q-1}\}c$. Once the equation was satisfied, its corresponding $\{v_0,\ldots,v_{q-1}\}c$ possesses a certain quadrature degree, fulfilling QES with a specific order.   
In classical approaches\mycite{Shan-2088,Shim-2780}, it is obtained through manually computing QES, which needs to construct the QES and separate the co-solving of $c$ and $\{\text{w}_\alpha\}$. 
Now as  the previous discussion shows that HCES is an equation system equivalent to QES but without involving $\{\text{w}_\alpha\}$, this relation can be directly constructed by calculating its Hermite polynomial coefficients $\{A_i\}$ in abscissa polynomial. 
Given a predefined $\{v_0,\ldots,v_{q-1}\}$ with an unknown lattice constant $c$, we substitute it into the abscissa polynomial with relation $\xi_\alpha = v_\alpha c$, and expand the product,
\begin{equation}\label{eq:abPl}
{W_q}\left( \xi  \right) = \prod\limits_{\alpha  = 0}^{q - 1} {\left( {\xi  - {v_\alpha }c} \right)}=\sum_{k=0}^{q}b_k\xi^k,
\end{equation}
where $b_k$ is an univariate polynomial of $c$. Introducing the explicit expressions for monomial in terms of Hermite polynomials, 
\begin{equation}\label{eq:HerExp}
{\xi ^k} = \frac{{k!}}{{{2^k}}}\sum\limits_{l=0}^{\left\lfloor {k/2} \right\rfloor } {\frac{1}{{l!\left( {k - 2l} \right)!}}{H_{k - 2l}}\left( \xi  \right)} ,
\end{equation}
where $\lfloor\cdot\rfloor$ is the floor function, \myeqref{eq:abPl} can be converted  into Hermite polynomial form,
\begin{equation}
\label{eq:HerForm}
{W_{q}}\left( \xi  \right) = \sum\limits_{i = 0}^{q} {{A_i}{H_i}\left( \xi  \right)} .
\end{equation}
Since \myeqref{eq:HerExp} does not involve new unknown variables, coefficient $A_i$ is still an univariate polynomial of $c$. \rvs{According to the pGHQ theory,  a series of $q\sim K^{th}$ HCES could be constructed for $\{v_0,\ldots,v_{q-1}\}c$, where $0\le K \le q-1$.
They cover all possible quadrature degrees of the discrete velocity set, from $q$ to $2q-1$. }
These series of HCES are the target univariate polynomial equation systems of $c$ which in classical approaches are constructed through solving QES. 
Hence the on-node lattices construction in the pGHQ scheme is simply performed on the abscissa polynomial without calculating QES and separating the co-solving of lattice constant $c$ and weights $\left\{\text{w}_\alpha\right\}$. 
Taking $\{0,\pm1,\pm5\}$ as an example, \rvs{after converting its abscissa polynomial into Hermite polynomial form,
\begin{equation}
{W_q}\left( \xi  \right) =\prod\limits_{\alpha  = 0}^{4} {\left( {\xi  - {v_\alpha }c} \right)} = {\xi ^5} - 26{c^2}{\xi ^3} + 25{c^4}{\xi ^1} = \sum\limits_{i = 0}^{5} {{A_i}{H_i}\left( \xi  \right)},
\end{equation}
where the coefficients $\{A_i\}$ read,
\begin{equation}\label{eq:Aexp}
{A_5} = \frac{1}{{32}},
{A_4} = 0,
{A_3} = \frac{5}{8} - \frac{{26{c^2}}}{8},
{A_2} = 0,
{A_1} = \frac{{15}}{8} - \frac{{78{c^2}}}{4} + \frac{{25{c^4}}}{2},
{A_0} = 0.
\end{equation}
its series of HCES could be directly generated. For an instance, its $5\sim 4^{th}$ HCES $\{A_i=0|i\le 4\}$ is,
\begin{equation}\label{eq:54HC}
 0=0, \frac{{15}}{8} - \frac{{78{c^2}}}{4} + \frac{{25{c^4}}}{2}=0, 0=0, \frac{5}{8} - \frac{{26{c^2}}}{8}=0,0=0
\end{equation}
Once this HCES has real-valued solution $c$, $\{0,\pm1,\pm5\}c$ satisfies $9$th QES. It worths noting that the $\{0,\pm1,\pm5\}c$ corresponding to \myeqref{eq:54HC} is the $5$th Gauss-Hermite quadrature, which as mentioned before is off-node. This off-node characteristic is reflected as no real solution $c$ for its HCES. 
\myeqref{eq:Aexp} presents the most significant advantage of the pGHQ scheme, i.e.
comparing with the generation of a specific univariate polynomial equation in classical approaches\mycite{Shan-2088,Shim-2780}, the pGHQ scheme systematically offers a series of HCES for lattice constant $c$ once  the expressions of coefficients $\{A_i\}$ were obtained.}

In practice, given a discrete velocity set $\{v_0,\ldots,v_{q-1}\}$, the quadrature degree of $\{v_0,\ldots,v_{q-1}\}c$ is required as high as possible so that it can be used to construct higher moment degree ED discretization. 
Therefore, one can start with solving its $q\sim (q-1)^{th}$ HCES where $K=q-1$ is the theoretically largest. Once this HCES had no real-valued solutions for $c$, one decreases $K$ by 1. As the construction of HCES shows, this decreasing is actually loosing the constraints on lattice constant $c$ by reducing the governing equations $\{A_i=0\}$.  This procedure is repeated until a real-valued $c$ is found. Its corresponding $K$ gives the quadrature degree of $\{v_0,\ldots,v_{q-1}\}c$, $q+K$, which indicates that this set can be used to construct the $u^{\left\lfloor {\left( {q + K} \right)/2} \right\rfloor }$ ED discretization. The construction of $f_\alpha^{eq}$ is illustrated in \myeqref{eq:disEq}. 
We designate this approach as the pGHQ scheme. 
It should be noted that there is no limitation on the given discrete velocity set. Given any kind of discrete velocity set, whether it is symmetric or asymmetric, the coefficients $\{A_i\}$ can always be obtained, and their procedures are unified with same formulas \myeqref{eq:abPl}$\sim$\myeqref{eq:HerForm}, which is another great advantage of the pGHQ scheme.  Hence the pGHQ scheme supports constructing all kinds of lattices, symmetric or asymmetric.

Comparing with the classical approaches\mycite{Shim-2780,Shan-2582}, the construction of univariate polynomial equation for lattice constant $c$ in the pGHQ scheme is systematical and general, supporting symmetric and asymmetric lattices and covering all quadrature degree. The procedure is concise without involving co-solving of $c$ and $\{\text{w}_\alpha\}$. And it can be mathematically proven that the univariate polynomial equation of $c$ in \myRefInd{Shim-2780,Shan-2582} equals HCES. Here,  a justification for the Shan scheme\mycite{Shan-2582} is offered in Sec.~\Rmnum{3} of the supporting information. It also worths noting that though pGHQ is proposed for the Hermite polynomial expansion theory, it also can be extended into entropic LB model. Actually it is the mathematical mechanism of a popular entropic LB discretization, the Karlin-Asinari scheme\mycite{KarlinAsinari-2430}. One can refer to the Sec.~\Rmnum{4} in the supporting information for the detail justification. 
This explains the interesting question\mycite{Shan-2582} that why for a given discrete velocity set one got the same lattice constant and weights under different schemes even under different theories.

\textit{Application.--} Since the pGHQ scheme offers a series of HCES covering full-range quadrature degree and supports all kinds of lattices,  a direct application is to construct optimal lattices, which restores the same moment degree with smallest discrete velocity set. 
In terms of the pGHQ scheme, given a $n$th-order moment degree on-node ED discretization, it is to construct a discrete velocity set $\{v_0,\ldots,v_{q-1}\}$ with smallest $q$, whose $q\sim\left( {2n-q}\right)^{th}$ HCES has real-valued solution for lattice constant $c$. The theoretically smallest number for $q$ is $\left(n+1\right)$, which indicates that its corresponding abscissa polynomial can be expressed as
\begin{equation}\label{eq:opHer}
{W_{\left(n+1\right)}}\left( \xi  \right) = \frac{1}{{{2^{n+1}}}}{H_{n+1}}\left( \xi  \right) + {A_{n}}{H_{n}}\left( \xi  \right).
\end{equation}
Unfortunately, the mechanism of tuning the coefficient $A_{n}$ to generate desirable zeros, which can fit into nodes, is not clear. Hence, the global optimal lattices is not available right now. However, since the procedure of the pGHQ scheme is unified for both symmetric and asymmetric and the core computation is solving HCES which is a univariate polynomial equation system, the pGHQ scheme is extremely suitable for computers. Therefore, limiting the range of the discrete velocity, a brute-force approach is available, which is enumerating all the possible discrete velocity set and identifying their feasibilities.
Here we search the local optimal lattices on $\left[ { - 10,10} \right]$ for $\left\{ {n = 3,4,5,6,7} \right\}$ moment degree ED discretization. The detailed procedures of searching local optimal lattices on $\left[ { - m,m} \right]$ for $u^{n}$ ED discretization are:
\begin{enumerate}[label={\bfseries \alph*)}]
	\item Set up $q$, start up with theoretically smallest number $q=n+1$
	\item\label{it:Enum} Enumerate all possible $q$-point discrete velocity sets on  $\left[ { - m,m} \right]$
	\item Solve $q\sim\left({2n-q}\right)^{th}$ HCES for each enumerated discrete velocity set. Identify the set with real-valued $c$ as a feasible lattices.
	\item\label{it:clos} All identified feasible sets are local optimal lattices  on  $\left[ { - m,m} \right]$. If there is no feasible lattices in the enumerated sets, increase $q$ by 1, repeat Step \ref{it:Enum} $\sim$ Step \ref{it:clos}
\end{enumerate}

Our result turns out that all these local optimal lattices keep the symmetric form,  $\left\{0,\pm v_1,..., \pm v_{n-1}\right\}$. The local optimal abscissa number $q$ on $\left[ { - 10,10} \right]$ has the relationship $q=2n-1$ with the moment degree $n$.  In order to verify the feasibility of asymmetric lattices, we continue our search with an extra point. The search shows that for a given $n$ moment degree ED discretization, the available lattices are extremely abundant. Taking $n=3$ moment degree ED discretization as an instance,  on range $\left[ { -10,10} \right]$ there are 20 5-point lattices (local optimal lattices) whose discrete velocity set has the form $\left\{0,\pm v_1, \pm v_{2}\right\}$ and 34636 6-point lattices where most of them are asymmetric lattices. \mytabref{tab:satRe} lists the detailed statistics of our search. And as a detailed illustration of the local optimal lattices, \mytabref{tab:comSmlDemo} presents the most compact local optimal lattices on $\left[ { - 10,10} \right]$ for $\left\{ {n = 3,4,5,6,7} \right\}$ moment degree ED discretization, whose discrete velocity is as close as possible to 0. To given a specific display of the abundance of available lattices, \mytabref{tab:lstQua} and \mytabref{tab:noSymQua}  list several symmetric and asymmetric lattices of $n=3$ moment degree ED discretization respectively.

\begin{table*}[htbp]
	\caption{Statistics of available on-node lattices for $\left\{ {n = 3,4,5,6,7} \right\}$ moment degree ED discretization on interval $\left[ { -10,10} \right]$. The table lists the local optimal $q$, the total number of available local optimal lattices, the total number of available $\left(q+1\right)$-point lattices in columns named ``Local optimal $q$'', ``$q$-point lattices'' and ``$\left(q+1\right)$-point lattices'' respectively. }\label{tab:satRe}
	\centering
	\begin{tabular}{p{1.5cm}p{2.6cm}*{2}{p{2.0cm}}}
		\hline\hline
		\noalign{\smallskip}
		Moment& Local optimal & $q$-point  & $(q+1)$-point\\
		degree $u^{n}$ & $q$ &  lattices & lattices \\
		\noalign{\smallskip}
		\hline
		\noalign{\smallskip}
		3     & 5     & 20    & 34636 \\
		4     & 7     & 120   & 138715 \\
		5     & 9     & 112   & 244218 \\
		6     & 11    & 252   & 211863 \\
		7     & 13    & 112   & 82684 \\
		\noalign{\smallskip}
		\hline\hline
	\end{tabular}%
	\centering
	\caption{Most compact local optimal on-node lattices on $\left[ { - 10,10} \right]$ for $\left\{ {n = 3,4,5,6,7} \right\}$ moment degree ED discretization }\label{tab:comSmlDemo}
		\resizebox{\textwidth}{!}{%
		\begin{tabular}{l l l l}
			\hline\hline
			\noalign{\smallskip}
			Moment&\multicolumn{1}{l}{\multirow{2}{*}{Lattices $\{v_0, \pm v_1, \cdots\}$}} & \multicolumn{1}{l}{Lattice} & \multicolumn{1}{l}{\multirow{2}{*}{weights $\left\{\text{w}_0, \text{w}_1, \cdots \right\}$}} \\
			 degree $u^n$&\multicolumn{1}{c}{} & \multicolumn{1}{l}{constant $c$} &  \\
			\noalign{\smallskip}
			\hline
			\noalign{\smallskip}
			 \multirow{2}{*}{3}&\multirow{2}{*}{$\{0,\pm1,\pm3\}$} & 1.1664E-00& \{6.3665E-01, 1.8141E-01, 2.6196E-04\}  \\
			   &   & 5.5343E-01& \{7.4464E-02, 4.1859E-01, 4.4182E-02\} \\
			\hline
			 4&$\{0,\pm1,\pm2,\pm3\}$ & 8.4639E-01& \{4.7667E-01, 2.3391E-01, 2.6938E-02, 8.1213E-04\} \\
			\hline
			 \multirow{2}{*}{5}&\multirow{2}{*}{$\{0,\pm1,\pm2,\pm3,\pm5\}$} & 8.1321E-01& \{4.5814E-01, 2.3734E-01, 3.2325E-02, 1.2641E-03, 8.9773E-07\}  \\
			   &   & 4.7942E-01& \{1.6724E-01, 3.0315E-01, 5.3303E-02, 5.7922E-02, 2.0013E-03\} \\
			\hline
			 6&$\{0,\pm1,\pm2,\pm3,\pm4,\pm5\}$ &6.8590E-01& \{3.8694E-01, 2.4178E-01, 5.8922E-02, 5.6153E-03, 2.0652E-04, 3.2745E-06\} \\
			\hline
			 \multirow{2}{*}{7}&\multirow{2}{*}{$\{0,\pm1,\pm2,\pm3,\pm4,\pm5,\pm7\}$} & 6.6344E-01 & \{3.7428E-01, 2.4105E-01, 6.4343E-02, 7.1316E-03, 3.2523E-04, 6.6163E-06, 3.0509E-09\}  \\
			  &   & 4.3240E-01& \{2.0928E-01, 2.3312E-01, 9.4051E-02, 5.6923E-02, 7.5008E-03, 3.7006E-03, 6.0784E-05\} \\
			\noalign{\smallskip}
			\hline\hline
	\end{tabular}}%
\end{table*}
\begin{table*}
	\caption{All available local optimal on-node lattices for $n=3$ moment degree ED discretization on interval $\left[-5,5\right]$. All feasible discrete velocity sets keep the symmetric form, $\left\{0,\pm v_1, \pm v_2\right\}$. $v_\alpha$ and $-v_\alpha$ share the same weight ${\text w}_\alpha$. Each feasible lattices has two lattice constants. For the sake of space saving, the two lattice constants $c$ and their corresponding weights $\left\{{\text w}_0,{\text w}_1,{\text w}_2\right\}$ are denoted by the symbols $\mp$ and $\pm$. The rational form is kept for comparing with existing ED discretizations.
	}\label{tab:lstQua}
	\centering
	\resizebox{\textwidth}{!}{%
		\begin{tabular}{ *{6}{l}}
			\hline\hline
			\noalign{\smallskip}
			\multicolumn{1}{c}{$\left\{0,\pm v_1,\pm v_2\right\}$} &\multicolumn{1}{c}{$c$}  &\multicolumn{1}{c}{${\text w}_0$} & \multicolumn{1}{c}{${\text w}_1$} & \multicolumn{1}{c}{${\text w}_2$}  \\
			\noalign{\smallskip}
			\hline
			\noalign{\smallskip}
			 $\left\{ {0, \pm 1, \pm 3} \right\} $ & $\sqrt {\left( {5 \mp \sqrt {10} } \right)} /{\sqrt 6 } $ & ${4}\left( {4 \mp \sqrt {10} } \right) /{45} $ & ${3}\left( {8 \pm \sqrt {10} } \right) /{80} $ & $\left( {16 \pm 5\sqrt {10} } \right) /{720} $ \\
			 $\left\{ {0, \pm 1, \pm 4} \right\} $ & $\sqrt {51 \mp \sqrt {1641} } /{8} $ & $\left( {93 \mp 17\sqrt {1641} } \right) /{1200} $ & $\left( {1959 \pm 29\sqrt {1641} } \right) /{4500} $ & $\left( {933 \pm 23\sqrt {1641} } \right) /{36000} $ \\
			$\left\{ {0, \pm 1, \pm 5} \right\} $ & $\sqrt {39 \mp \sqrt {1146} } /{5\sqrt 2 } $ & $\left( { - 528 \mp 52\sqrt {1146} } \right) /{1875} $ & $\left( {2208 \pm 47\sqrt {1146} } \right) /{3600} $ & $\left( {2472 \pm 73\sqrt {1146} } \right) /{90000} $ \\
			 $\left\{ {0, \pm 2, \pm 5} \right\} $ & $\sqrt {87 \mp \sqrt {1569} } /{20} $ & $\left( {3477 \mp 29\sqrt {1569} } \right) /{7500} $ & $\left( {3153 \pm 19\sqrt {1569} } \right) /{12600} $ & $\left( {2829 \pm 67\sqrt {1569} } \right) /{157500} $ \\ 
			\noalign{\smallskip}
			\hline\hline
	\end{tabular}}
	\caption{Available asymmetric on-node lattices for $n=3$ moment degree ED discretization. }\label{tab:noSymQua}
	\centering
	\begin{tabular}{ l | @{~}c @{~}|*{6}{@{~}c @{~}}}
		\hline\hline
		\noalign{\smallskip}
	    \multicolumn{1}{c}{$\left\{v_0,v_1,...,v_5\right\}$}    & \multicolumn{1}{c}{$c$} & \multicolumn{6}{c}{$\left\{{\text w}_0,{\text w}_1,...{\text w}_5\right\}$} \\
		\noalign{\smallskip}
		\hline
		\noalign{\smallskip}
		 $\{-5, -2, -1, 1, 2, 4\}$ &0.381641&0.019568&0.302751&0.094520&0.505439&0.009237&0.068487\\
		 $\{-4, -3, -1, 1, 2, 4\}$ &0.450877&0.016717&0.054744&0.451349&0.323613&0.127736&0.025841\\
		 $\{-3, -1, 0, 1, 2, 4\}$ &0.521696&0.059199&0.366034&0.198867&0.227904&0.138130&0.009866\\
		 $\{-3, -1, 0, 1, 2, 5\}$ &0.553432&0.076212&0.294489&0.352957&0.040448&0.232265&0.003629\\
		\noalign{\smallskip}
		\hline\hline
	\end{tabular}%
\end{table*}

\begin{figure}[htbp]
	\centering
	\includegraphics[width=0.6\textwidth]{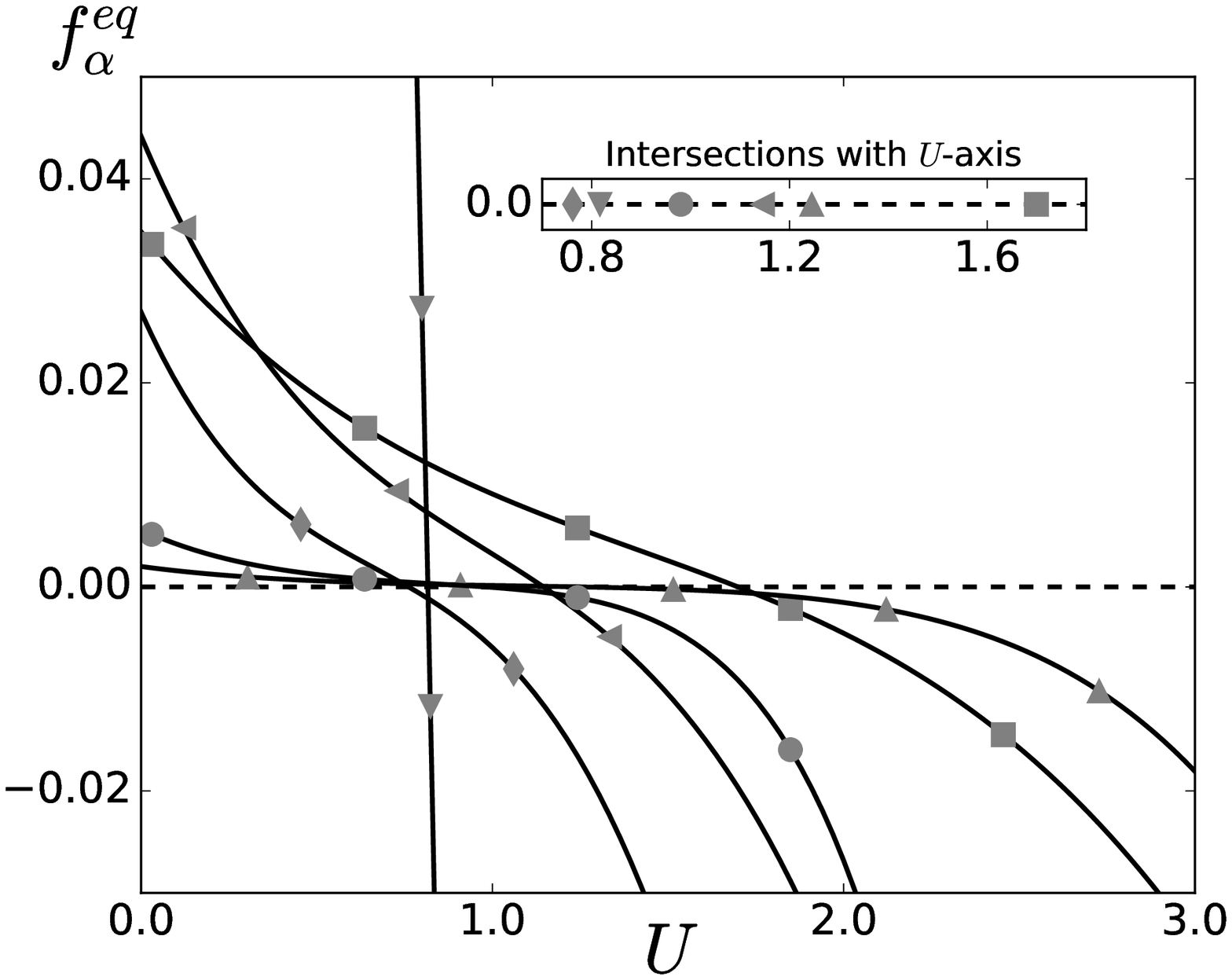}
	\caption{The profiles of first-going-negative equilibrium distributions $f_{\alpha}^{eq}$ as functions of $U$. The figure only renders the positive $U$-axis. Since all lattices in the figure are symmetric, the positivity on the negative $U$-axis is the same though the corresponding $v_{\alpha}$ turns to $-v_{\alpha}$.
		The profiles contain: 
		{\bfseries a).} line with symbol $\blacktriangledown$ is for $f_{\alpha}^{eq}$ with $v_{\alpha}=0$ and $c=1/\sqrt{2RT}=1.2247$ which, as $U$ increases, first goes negative in all equilibrium distributions of $\{0,\pm1\}$, 
		{\bfseries b).} $\blacksquare$ for $v_{\alpha}=-5$ and $c=0.3442$ in $\{0,\pm2, \pm5\}$,  
		{\bfseries c).} $\blacktriangleleft$ for $v_{\alpha}=-3$ and $c=0.5534$ in $\{0,\pm1,\pm3\}$,  
		{\bfseries d).} $\blacklozenge$ for $v_{\alpha}=-2$ and $c=0.8464$ in $\{0,\pm1,\pm2,\pm3\}$,  
		{\bfseries e).} $\blacktriangle$ for $v_{\alpha}=-5$ and $c=0.4794$ in $\{0,\pm1,\pm2,\pm3,\pm5\}$,  
		{\bfseries f).} $\CIRCLE$ for $v_{\alpha}=-3$ and $c=0.6859$ in $\{0,\pm1,\pm2,\pm3,\pm4,\pm5\}$.
		The inner panel renders their intersections with $U$-axis, above which the $ f_{\alpha}^{eq}$ will become negative. The specific values of intersections for \{{\bfseries a}, {\bfseries b}, {\bfseries c}, {\bfseries d}, {\bfseries e}, {\bfseries f}\} are $\sim$\{$0.82$, $1.70$, $1.15$, $0.76$, $1.25$, $0.98$\}.
		Since the plotted  $f_{\alpha}^{eq}$ are the first-going-negative equilibrium distributions, then the inner panel demonstrates the lattices positivity range of $U$.}\label{fig:anaFe}
\end{figure}

\begin{figure}[htbp]
	\centering
	\includegraphics[width=0.6\textwidth]{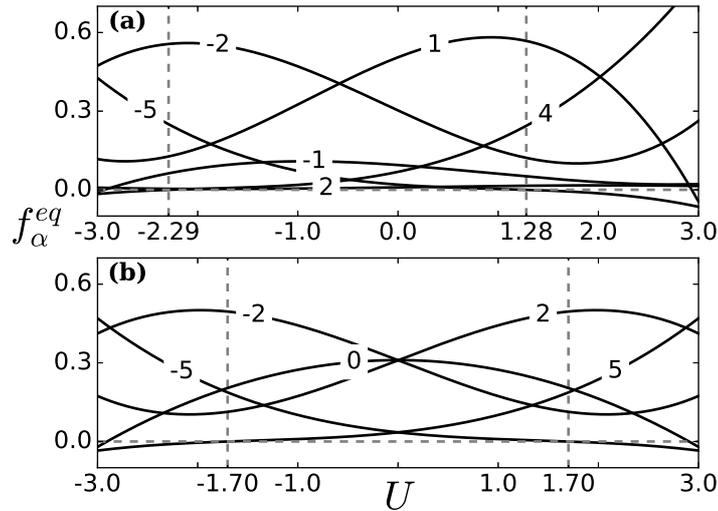}
	\caption{The profiles of equilibrium distributions $f_{\alpha}^{eq}$ as functions of $U$: {\bfseries a).} lattices $\{-5, -2, -1, 1, 2, 4\}$ with lattice constant $c=0.3816$; {\bfseries b).} lattices $\{0, \pm2, \pm5\}$ with lattice constant $c=0.3442$. The label on a curve is its corresponding $v_\alpha$. The intervals of $U$ between vertical dash lines are lattices positivity ranges.}\label{fig:nonSymmFeq}
\end{figure}

\textit{Implication.--} A direct implication of lattices abundance is its impact on the positivity of equilibrium distributions, i.e. the range of macro velocity on which all equilibrium distributions remain positive. As a negative equilibrium distribution violates the physical nature of particle kinetics, the positivity is considered as one major factor to the numerical stability of the LB method. 
We analyzed lattices $\{0,\pm1\}$, $\{0,\pm2, \pm5\}$, $\{0,\pm1,\pm3\}$, $\{0,\pm1,\pm2,\pm3\}$, $\{0,\pm1,\pm2,\pm3,\pm5\}$, $\{0,\pm1,\pm2,\pm3,\pm4,\pm5\}$. 
For lattices $\{0,\pm2, \pm5\}$, $\{0,\pm1,\pm3\}$, $\{0,\pm1,\pm2,\pm3,\pm5\}$ have two feasible lattice constants $c$, we take the $c$ with a wider positivity. The analysis shows that lattices $\{0,\pm2, \pm5\}$ has the widest positivity though its retained moment degree is only $u^3$. Meanwhile the positivity of highest retaining-moment-degree lattices $\{0,\pm1,\pm2,\pm3,\pm4,\pm5\}$ is merely better than $\{0,\pm1\}$, $\{0,\pm1,\pm2,\pm3\}$. To demonstrate it, Fig.~\ref{fig:anaFe} plots their equilibrium distributions which firstly go negative as the macro velocity increases.
The asymmetric lattices also demonstrates its capability on modifying the positivity on a specific range of $U$. Fig.~\ref{fig:nonSymmFeq} plots a comparison of lattices $\{-5, -2, -1, 1, 2, 4\}$ and $\{0, \pm2, \pm5\}$. It shows that lattices $\{-5, -2, -1, 1, 2, 4\}$ shifts the positivity range of $\{0, \pm2, \pm5\}$ left with approximatively $-0.5$ on $U$-axis.
The analysis indicates that the discrete velocity set could be a significant impact to the numerical stability of LB method. It offers a direction to improve LB numerical stability. And our identified lattices offer a database for the further study. For detailed investigation, it is beyond the scope of this paper and shall be addressed in a separate publication.

\textit{Conclusion.--} We propose a new mathematical tool, pGHQ, to construct on-node LB lattices under a Cartesian coordinate system in this paper. To the best of our knowledge, it is the first time to derive and employ this mathematical tool in the context of LB method.   
pGHQ is general. It can be extended into the entropic LB model though firstly proposed for the Hermite polynomial expansion theory. 
The pGHQ scheme avoids the tedious QES solving. Comparing with the \mrvs{existing} classical approaches, our scheme has the following advantages: \textbf{a).} the algorithm is extremely concise, \textbf{b).}the procedure of constructing univariate polynomial equations is unified for both symmetric and asymmetric lattices, \textbf{c).} the generated univariate polynomial equation system covers full-range quadrature degree of the given $\{v_0,\ldots,v_{q-1}\}$. 
We employ the pGHQ scheme to search the local optimal and asymmetric lattices on $\left[ { - 10,10} \right]$ for $\left\{ {n = 3,4,5,6,7} \right\}$ moment degree ED discretization. The search reveals a surprising abundance of available lattices. Our brief analysis shows that the discrete velocity set is significant to the positivity of equilibrium distribution, which is one major impact to the numerical stability of LB method. Hence the results of the pGHQ scheme lay a foundation for further investigation on improving the numerical stability of LB method by modifying the discrete velocity set.

\addcontentsline{toc}{chapter}{Acknowledgment}
\section*{Acknowledgment}
Huanfeng Ye would like to express his gratitude to Dr. Yung-an Chao for helpful discussion on the paper organization.

\addcontentsline{toc}{chapter}{References}

%

\end{document}